\documentclass[12pt]{article}
\pdfoutput=1
\usepackage{amsfonts,amsmath,amssymb}
\usepackage[paper=letterpaper,margin=1.0in]{geometry}
\usepackage{graphicx}
\usepackage{cite}
\usepackage{chngcntr}
\usepackage{tocloft}
\usepackage{hyperref}
\usepackage{xcolor}
\usepackage{mathtools}

\usepackage[sort,numbers]{natbib}

\newcommand\be{\begin{equation}}
\newcommand\ee{\end{equation}}
\newcommand\bea{\begin{eqnarray}}
\newcommand\eea{\end{eqnarray}}
\newcommand\ba{\begin{array}}
\newcommand\ea{\end{array}}
\newcommand\ben{\begin{enumerate}}
\newcommand\een{\end{enumerate}}
\newcommand\bi{\begin{itemize}}
\newcommand\ei{\end{itemize}}
\newcommand\bc{\begin{center}}
\newcommand\ec{\end{center}}
\newcommand\bfig{\begin{figure}}
\newcommand\efig{\end{figure}}
\newcommand\bq{\begin{quotation}}
\newcommand\eq{\end{quotation}}
\newcommand\bt{\begin{table}}
\newcommand\et{\end{table}}
\newcommand\btab{\begin{tabular}}
\newcommand\etab{\end{tabular}}

\newcommand\tr{\mathrm{Tr}}

\counterwithin*{equation}{section}
\renewcommand\thefootnote{\fnsymbol{footnote}}
\newcommand\comment[1]{}

\renewcommand\tilde{\widetilde}

\begin{document}

${}$

\vspace{.5cm}

\begin{center}
{\Huge \textbf{Spinning probes and helices in AdS$_3$ }}

\vspace{.5cm}

Piermarco Fonda\footnote{\href{mailto:fonda@lorentz.leidenuniv.nl}{\texttt{fonda@lorentz.leidenuniv.nl}}}$^{1}$,
Diego Liska\footnote{\href{mailto:lis14392@uvg.edu.gt}{\texttt{lis14392@uvg.edu.gt}}}$^2$,
\'Alvaro V\'eliz-Osorio\footnote{\href{mailto:avelizosorio@th.if.uj.edu.pl}{\texttt{avelizosorio@th.if.uj.edu.pl}}}$^{3}$

\vspace{.33cm}

\vspace*{.2cm}
{${}^{2}$ Instituut Lorentz, Universiteit Leiden,\\ P.O. Box 9506, 2300 RA Leiden, The Netherlands\\}

\vspace*{.2cm}
{${}^{2}$ Department of Physics, Universidad del Valle de Guatemala,\\
18 Avenida 11-95, Zona 15 Guatemala, Guatemala}

\vspace*{.2cm}
{${}^{3}$ M. Smoluchowski Institute of Physics, Jagiellonian University,\\
\L{}ojasiewicza 11, 30-348 Krak\'ow, Poland\\}

\end{center}

\vspace{1cm}

\centerline{\textbf{Abstract}}
\bigskip
We study extremal curves associated with a functional which is linear in the curve's torsion. The functional in question is known to capture the properties of entanglement entropy for two-dimensional conformal field theories with chiral anomalies and has potential applications in elucidating the equilibrium shape of elastic linear structures. We derive the equations that determine the shape of its extremal curves in general ambient spaces in terms of geometric quantities. We show that the solutions to these shape equations correspond to a three-dimensional version of Mathisson's helical motions for the centers of mass of spinning probes. Thereafter, we focus on the case of maximally symmetric spaces, where solutions correspond to cylindrical helices and find that the Lancret ratio of these  equals the relative speed between the Mathisson-Pirani and the Tulczyjew-Dixon observers. Finally, we construct all possible helical motions in three-dimensional manifolds with constant negative curvature. In particular, we discover a rich space of helices in AdS$_3$ which we explore in detail.
\newpage

\setcounter{tocdepth}{2}
\tableofcontents

\setcounter{footnote}{0}
\renewcommand\thefootnote{\arabic{footnote}}

\section{Introduction}\label{sec:intro}

This work is devoted to the study of helical curves in three-dimensional manifolds. These are curves $\gamma$ which extremize the geometric functional
\be  
{\cal F}[\gamma] =\mathfrak{m}\, \ell(\gamma) + \mathfrak{s} \int_\gamma \tau\,,
\label{torsionaction1}
\ee
where $\ell(\gamma)$ is the curve's length and $\tau$ is its extrinsic torsion. Intuitively, $\tau$  is a measure of \textit{non-planarity} of the curve. This functional is rich in geometrical structure and applications: for space-like curves embedded in asymptotically Anti-de Sitter spaces, it encodes the holographic entanglement entropy of two-dimensional conformal field theories with chiral anomalies \cite{Castro:2014tta}. Furthermore, its extrema delineate the time-like trajectories of the center of mass of spinning particles in three-dimensional curved spacetimes. Remarkably, its applications are not limited to holography and general relativity since functionals of this kind can be used to understand the behaviour of elastic linear structures such as proteins and polymers \cite{doi:10.1093/qjmam/hbn012}\cite{1751-8121-47-35-355201}.

In a ground-breaking insight, Ryu and Takayanagi \cite{Ryu:2006bv} observed that the entanglement entropy for two-dimensional conformal field theories can be computed by evaluating the length functional for curves in AdS$_3$ on its extrema, which correspond to geodesics. This prescription is valid only for field theories whose holographic dual is Einstein gravity. Over the years,  a number of generalizations to this prescription have been developed in order to accommodate gravity duals endowed with higher-curvature corrections. The new entanglement functionals involve terms containing geometric objects such as the extrinsic curvature and projections of the ambient curvature \cite{Bhattacharyya:2013gra,Dong:2013qoa,Camps:2013zua}. Beyond geodesics, there is a rich space of extrema for these functionals  even in the simplest cases \cite{Fonda:2016ine}. A generalization of the Ryu-Takayanagi prescription of particular relevance for the present discussion is the entanglement functional associated with two-dimensional field theories suffering from gravitational anomalies. In two dimensional conformal field theories this anomaly manifests itself as a discrepancy between the left and right central charges. Provided such theory admits a gravitational dual, this would necessarily contain a gravitational Chern-Simons term. Based on this fact, it was discovered in \cite{Castro:2014tta} that holographic entanglement entropy must be computed using the functional \eqref{torsionaction1}. The anomalous nature of the theory is succinctly captured by the torsion term and geodesics must be replaced by extrema of \eqref{torsionaction1}.

 Such curves are closely related to the trajectories of free falling spinning probes in curved spacetimes. The motion of these probes is known to be governed by the Mathisson-Papapetrou-Dixon equations \cite{Mathisson:1937zz, Papapetrou:1951pa, Dixon:1970zz}. These equations are undetermined up to a spin supplementary condition. This undeterminacy is a manifestation of the fact that in order to track the motion of an extended body using a single curve one must choose a reference point in the body. Newtonian intuition would suggest that one must select this point to be the center of mass of the body. However, the notion of center of mass is observer-dependent in relativistic systems: in fact, it is the job of the spin supplementary condition to single out a specific observer. Many different supplementary conditions have been considered in the literature. Arguably, the better known ones are the Tulczejew-Dixon (TD) and the Mathisson-Pirani (MP) conditions; the former selects an observer in the zero-momentum frame, while the latter opts for an observer comoving with the center of mass as seen by herself. The MP condition is known to give rise to helical motions \cite{Mathisson:1937} which were regarded as unphysical for a long time due to a subtle misapprehension. The nature of these \emph{Mathisson's helical motions} has been settled only recently in \cite{Costa:2011zn, Costa:2017kdr} where it was shown that they are in fact physically sound solutions. As shown in this work, the extrema of the functional \eqref{torsionaction1} describe Mathisson's helical motions in three dimensions. The geometric formalism developed here allows to relate physical notions to important geometrical quantites. For instance it is shown that the relative speed between the TD and the MP observers corresponds to the curvature to torsion ratio also known as the \emph{Lancret ratio} of the helix.

This paper is organized as follows: In Section \ref{sec:shape equations} we present the basic geometrical setup and deduce the equations satisfied by the extrema of the functional \eqref{torsionaction1}. In Section \eqref{sec:spinning particles} we show the equivalence of the extrema of the functional \eqref{torsionaction1} and three-dimensional Mathisson's helical motions. In Section \ref{helices} we construct analitically all possible helical motions in $\mathbb{H}_3$ and AdS$_3$ and discuss their properties.

\section{Geometrical setup and shape equations}\label{sec:shape equations}

In this section we elaborate on the geometrical content of the functional \eqref{torsionaction1} and deduce the equations satisfied by its extrema. We discuss the general setup using the nomenclature of \cite{Fonda:2016ine} where the subject is developed more thoroughly. Consider the embedding of a $p$-dimensional manifold $\Sigma$ into a $d$-dimensional manifold $M$ ($p<d$) endowed with a (pseudo) Riemannian metric $g_{\mu\nu}$. Locally, we can write this map as $x^\mu(\sigma^i)$, where $\mu=1,\dots,d$ and $i=1,\dots,p$. Hereafter, we will refer to the embedded image of $\Sigma$ also as $\Sigma$ whenever this doesn't lead to misunderstandings. At each point $p\in\Sigma$ the tangent space $T_pM$ can be decomposed into the space of vectors tangent to $\Sigma$ and its orthogonal complement. The former corresponds to the span of 
\be\label{tangent}
t_i^{\mu}=\frac{\partial x^\mu}{\partial \sigma^i}=\partial_i x^\mu,
\ee
while the latter can be generated by vectors $n^A_\mu$, with $A=1,\dots, d-p$ chosen such that 
\be\label{normal}
n^A_\mu\, t_i^{\mu}=0\qquad \text{and}\qquad g_{\mu\nu}n_A^\mu n_B^\nu=\eta_{AB}\,,
\ee
where $\eta_{AB}$ is given by the matrix  $\eta_{AB}=\mathrm{diag}(-1,\dots,-1,1\dots,1)$. The number of negative eigenvalues in $\eta_{AB}$ depends on the signature of $g_{\mu\nu}$ and the nature of the embedding. In the following, we use mixed indices to denote projections of ambient quantites into tangent and normal directions, for instance:
\be
R^{AiC}_{\;\;\;\;\;\;\;\;j }=R_{\mu\nu\rho\sigma}n^{\mu\, A}t^{\nu\, i}n^{\rho\, C} t^{\sigma}_j\,.
\ee
 Notice that the vectors $n^A_\mu$  are defined up to transformations 
\be\label{gauge1}
 n^A_\mu\rightarrow {\cal M}^A_{\;\;C}\left(\sigma^i\right)\, n^C_\mu\qquad\text{such that}\qquad {\cal M}^A_{\;\;C}{\cal M}^B_{\;\;D}\, \eta_{AB}=\eta_{CD}\,.
\ee
This ambiguity gives rise to a gauge theory on the normal bundle, which for a Riemannian ambient space corresponds to an $SO(d-p)$ gauge theory.

The geometrical properties of the embedding of $\Sigma\hookrightarrow M$ can be divided into \emph{intrinsic} and \emph{extrinsic}. Intrinsic properties are those obtained from the induced metric (or first fundamental form) 
\be\label{induced}
h_{ij}= g_{\mu\nu} t_i^{\mu} t_j^{\nu}\,.
\ee
We can associate a Levi-Civita connection $\tilde\nabla_i$ to $h_{ij}$ and hence obtain the intrinsic Riemman tensor ${\cal R}_{ijkl}$ and its relevant contractions.
In turn, the extrinsic properties can be found by studying how the tangent and normal vectors change as we move along $\Sigma$. We can decompose the directional derivatives $t^\mu_i\nabla_\mu={\cal D}_i$ of the normal vectors into their tangent and normal contributions as:
\begin{align}
{\cal D}_i\, n^{A\mu}=K_{ij}^A\,t^{j\mu}-T^{AB}_in^\mu_B \, . \label{directional2}
\end{align} 
The coefficients of the above decomposition define the key objects that encode the extrinsic geometry, the \emph{extrinsic curvatures}
\be\label{extrinsic curvature}
K_{ij}^A=t_j^\mu\,{\cal D}_i\, n^{A}_\mu\,,
\ee
and the \emph{extrinsic torsions}
\be\label{extrinsic torsion}
T^{AB}_i=n^{A}_\mu\,{\cal D}_i\, n^{B\mu}\,.
\ee
It is important to point out that ambient, intrinsic and extrinsic quantities partake in subtle relations such as the generalized Gauss identity
\be
{\cal R}_{ijkl}=R_{ijkl}+\eta_{AB}K^A_{[il}K^B_{jk]}\,,
\ee
which relates the extrinsic curvatures with the ambient and intrinsic Riemann tensors. For a detailed discussion on this subject  we refer the reader to \cite{Fonda:2016ine}.

It is natural to ask how do the extrinsic quantities depend upon the choice of normal frame. It can be shown that under the gauge transformations \eqref{gauge1} the extrinsic quantities transform as \cite{Capovilla:1994bs,Fonda:2016ine}
\be\label{gauge extrinsic curvature}
K_{ij}^A\rightarrow {\cal M}^A_{\;\;C}\left(\sigma^i\right) K_{ij}^C\,,
\ee
and
\be\label{gauge extrinsic torsion}
T^{AB}_i\rightarrow {\cal M}^{CA}\,\partial_i {\cal M}^D_{\;\;A}+  {\cal M}^C_{\;\;A} {\cal M}^D_{\;\;B}\, T^{AB}_i\,.
\ee
Notice that $T^{AB}_i$ transforms exactly as a gauge connection. Thus,  $T^{AB}_i$ can be used to construct a gauge covariant derivative in the normal bundle 
\be\label{gauge covariant}
\tilde D_{i\;B}^A V_j^B=\tilde\nabla_i V_j^A +T^{AB}_i\eta_{BC} V_j^C\,.
\ee
With the help of this connection it possible to define gauge theoretical objects in the normal bundle such as field strengths
 and, in the appropriate dimensions, Chern-Simons terms. This gauge symmetry provides a useful guiding principle for writing consistent geometric effective actions. A wide variety of physical situations can be described by functionals constructed with the geometric objects introduced above. For instance, the gauge invariant term  
\be
\sqrt{h}\, \tr\left(K^A\right)\tr\left(K_A\right)
\ee
in the Lagrangian plays an important role in the study of elasticity \cite{Capovilla:1994bs} as well as the computation of entanglement entropy via holography \cite{Bhattacharyya:2013gra,Dong:2013qoa,Camps:2013zua} depending on the ambient space considered.


In order to find the equations of motion corresponding to a geometric functional one must understand how geometric quantities respond to variations of the underlying sub-manifold of the form
\be\label{variation1}
x^\mu\rightarrow x^\mu+\delta x^\mu\,.
\ee
There are a number of subtleties associated with performing these variations, in the present work we just outline the procedure and invite the interested reader to consult \cite{Fonda:2016ine, Capovilla:1994bs} for detailed expositions.
The first step is to decompose the variation into its tangent and normal components
\be\label{variation2}
x^\mu\rightarrow x^\mu+\varepsilon^i\left(\sigma\right) t_i^\mu+\varepsilon^A\left(\sigma\right)n_A^\mu\,.
\ee
The advantage of this decomposition becomes clear once we notice that the tangential variations correspond to reparametrisations of $\Sigma$. Hence, tangential variations produce total derivatives which can contribute only as boundary terms. To find the equations of motion associated to the functional \eqref{torsionaction1} we just need the variations of $h_{ij}$ and $T_i^{AB}$. These normal variations can be obtained by taking Lie derivatives along the vector field ${n^\mu}=\varepsilon^A\left(\sigma\right)n_A^\mu$. The normal variation of the induced metric reads
\be\label{variation h}
{\cal L}_{n}\,h_{ij}=2\,\varepsilon_A K^A_{ij}\,,
\ee
which implies that 
\be\label{variation sqr h}
{\cal L}_{n}\,\sqrt{h}=\varepsilon_A \sqrt{h}\, \tr K^A\,.
\ee
 In turn, for the extrinsic torsion we find
\begin{align}\label{variation T}
{\cal L}_n T_i^{AB}=&-K^A_{ij}\left(\tilde\nabla_k \varepsilon^B-T_k^{BC}\varepsilon _C\right)h^{jk}+\varepsilon_D\,\Theta^{DA}_{\;\;\;\;\;C} T^{CB}_i-(A\leftrightarrow B) \nonumber\\
&+\tilde\nabla_j( \varepsilon_C\,\Theta^{CAB}) +R^{ABC}_{\;\;\;\;\;\;\;\;i }\varepsilon_C\,, 
\end{align}
where we introduced 
\be\label{theta}
\varepsilon_A\Theta^{ABC}=n^{\mu C}n^{\nu}\nabla_\nu n^B_\mu\,,
\ee
which is antisymmetric in the last two indices.

Now, we return to the case of interest namely curves ($p=1$) embedded in a three-manifold ($d=3$). Whenever we are working in codimension two we can define
\be\label{tau}
\tau_i = \frac{1}{2} \epsilon_{AB} T_i^{AB}\,,
\ee
with $\epsilon_{12}=-\epsilon_{21}=1$. In the case of cuves we omit the tangential index $i$ to avoid clutter, it is nevertheless important to be aware that $\tau$ is actually a one-form and not a scalar function. Depending on the signature of the normal frame the gauge group is either $U(1)$ or the group of squeeze mappings. Under these transformations, $\tau$ transforms as 
\be
\tau \to \tau  +\partial_i \psi \,,
\ee
where $\psi$ is the local angle parametrizing the local rotation of the normal frame. This means that the functional 
\be  
 \int^{\sigma_f}_{\sigma_i} d\sigma\, \tau  \,,
\ee
is not gauge-invariant, but rather receives boundary - in this case end-point - contributions 
\be
 \int^{\sigma_f}_{\sigma_i}  d\sigma \, \tau   \to  \int^{s_f}_{s_i}  d\sigma \, \tau  +\psi(\sigma_f)-\psi(\sigma_i)\,,
\ee
in a manner reminiscent of Chern-Simons theory. 

To obtain the equations of motion, we contract Eq.\,\eqref{variation T} with $\epsilon_{AB}$ and find that 
\begin{align}\label{normal variation tau}
\mathcal{L}_n \int_\Sigma ds \; \tau
=
 \int^{s_f}_{s_i}  ds\,
\varepsilon^A (s)\,
\left[
\epsilon_{AB}  \,
\tilde{D}^B_{s\;C} \tr K^C
-
\epsilon_{DE}  
R_{s\;\;A}^{\;\;D\;\;E} 
\right]
 \,,
\end{align}
up to boundary contributions.
Combining this result with \eqref{variation sqr h} implies that the extrema of the functional \eqref{torsionaction1} satisfy 
\be 
\mathfrak{m}\,\tr K_A  
+
\mathfrak{s} 
\left(  \epsilon_{AB}  
\tilde{D}^B_{s\;C}\tr K^C
-
\epsilon_{DE}  
R_{s\;\;A}^{\;\;D\;\;E}  \right) = 0\,.
\ee
To unburden the notation we write $k^A=\tr\, K^A $ and define $(\tilde{D} k)^B= \tilde{D}^B_{s\;C} k^C\,,$ so that the equation can be rewritten as
\be 
\mathfrak{m}\,k_A  
+
\mathfrak{s} \,  \epsilon_{AB}
\left[
(\tilde{D} k)^B
+
R_s^{\;B}\right] = 0\,,\label{shape torsion2}
\ee
where we also used the fact that in three dimensions the Riemann tensor can be expressed in terms of the Ricci tensor. 
We refer to Eqs.\,\eqref{shape torsion2} as the \emph{shape equations} and call its solutions \emph{extremal curves} or just \textit{extrema} in the following.
Observe that if we set $\mathfrak{s}=0$ the shape equations reduce to $k^A=0$, whose solutions are geodesics. It is our objective to explore further the possible of solutions of the shape equations.

To understand better the space of solutions of \eqref{shape torsion2} it is important to be able to associate gauge-invariant quantities to different extrema. An invariant of paramount importance is the \emph{total curvature} or \emph{Frenet-Serret (FS) curvature} defined by
\be\label{KFS}
k^2_{\mathrm{FS}}=\eta_{AB}k^A k^B\,.
\ee
The above, can also be regarded as the norm of the \emph{extrinsic curvature vector }
\be 
k^{\mu}=k^A n_A^{\mu}\,.
\ee
In Lorentzian spaces, this vector can be spacelike, timelike or null hence we must
bear in mind that $k^2_{\mathrm{FS}}$ might be negative. The behaviour of the FS curvature along the curve can be inferred from \eqref{shape torsion2} indeed, we find that
\be
\partial_s k^2_{\mathrm{FS}}=-2 R_s^{\;B}k_B \label{change in kFS}\,.
\ee
Therefore, whenever $R_s^{\;B}=0$ the Frenet-Serret curvature is constant along the curve. This is the case for Einstein spaces where the Ricci tensor is proportional to the metric tensor.

The shape equations \eqref{shape torsion2} transform covariantly under the gauge transformations \eqref{gauge1}, and we should fix a convenient gauge to solve them. There are two gauge choices which will play important roles in this work.
The first one is the Frenet-Serret gauge, which corresponds to a choice of normal vectors for which the extrinsic curvature in one of the normal directions vanishes identically. This means that all the curvature is manifested in the other normal direction, which we call FS normal, $ n_{\mathrm{FS}}^{\mu}$, such that
\be
\kappa_{\mathrm{FS}}=t_\mu\,{\cal D}_s\, n^\mu_{\mathrm{FS}}\,.
\ee
In Lorentzian spaces one must be particularly careful in selecting the FS frame: in fact, the FS normal must be endowed with the same causal nature of the vector $k^{\mu}$. Indeed, we have 
\be\label{k subtle}
 \kappa_{\mathrm{FS}}^2=\chi_{\mathrm{FS}}\, k^2_{\mathrm{FS}}\qquad \chi_{\mathrm{FS}}=g_{\mu\nu}  n^\mu_{\mathrm{FS}} n^\nu_{\mathrm{FS}}=\pm 1,0
\ee
which keeps track of whether $k^{\mu}$ is spacelike, timelike or null. An important consequence of this discussion is that for curves with varying $ k^2_{\mathrm{FS}}$ the choice of the FS frame must be revised whenever this quantity changes sign. The following example from \cite{2008arXiv0810.3351L} illustrates the point. Consider the curve 
\be
\gamma(s)=\left( \frac{1}{2}\left(s \sqrt{s^2-1}-\log(s+\sqrt{s^2-1})\right),\,\cos(s)+s \sin(s),\,\sin(s)-s\cos(s)\right)
\ee
 defined in the range $s\in (1,\infty)$ and embedded in flat Minkowski space $(t,x,y)$. For this curve, we have
\be
 \chi_{\mathrm{FS}}=\mathrm{sgn}\left(\frac{s^4-s^2-1}{s^2-1}\right)\,
\ee
and the causal structure of the FS normal vector can change as we vary $s$. It is important to avoid overlooking these subtleties when dealing with curves for which  $\dot k^2_{\mathrm{FS}}\neq 0$.

 The extrinsic geometry in a given frame is characterized by $k^A$ and $\tau$. Above, we explained how to obtain the curvatures in the FS frame starting from any given frame. Now, we must find the torsion $\tau_{\mathrm{FS}}$  in this gauge. We achieve this aim  using the invariant
\be
\epsilon_{AB}k^A(\tilde{D} k)^B\,, 
\ee
which can be used to show that 
\be\label{TFS}
k_{\mathrm{FS}}^2\left(\tau-\tau_{\mathrm{FS}}\right)=\epsilon_{AB}k^A \partial_s k^B\,.
\ee
Therefore, with the help of Eqs.\,\eqref{KFS}, \eqref{k subtle} and \eqref{TFS} it is possible to find the FS quantities of a curve starting from any given frame. 

The second convenient choice of gauge comes from demanding the torsion to be vanishing along the curve: there are never obstructions to this choice since the field strength associated to $\tau$ is always vanishing for one-dimensional curves. We refer to this gauge as the \emph{Fermi-Walker (FW) frame} since it can be used to to define non-inertial and non-rotating frames. We will revisit this point in section \eqref{sec:spinning particles}. We notice that the main advantage of this frame is the simplification of the gauge-covariant derivatives into ordinary derivatives.

\section{Extrema and spinning probes}\label{sec:spinning particles}

It is well known that in general relativity the trajectory of a test point particle is determined by the geodesic equations. This is the appropriate description for a particle with no internal structure. In the language of the previous section geodesic curves correspond to solutions of Eqs.\,\eqref{shape torsion2} with $\mathfrak{s}=0$. It is natural to wonder if more general solutions to these equations have an analogous interpretation in terms of test objects. Indeed, as announced by the title of this work, the extrema of \eqref{shape torsion2} can be associated with the trajectories of the centers of mass of spinning probes. More specifically, the solutions of the shape equations \eqref{shape torsion2} are closely related to \emph{Mathisson's helical motions} for spinning particles \cite{Mathisson:1937}. The nature of these trajectories has been a source of contention for many years and their true character has been elucidated only recently in \cite{Costa:2011zn}.

An extended relativistic body can be studied with the help of a multipole expansion performed about a suitably chosen reference worldline $\tilde x^\mu(s)$. At first order in this expansion, the motion of the body is encoded in two \emph{moments} of $T^{\mu\nu}$, the momentum $p^\mu$ and the angular momentum ${\cal S}^{\mu\nu}$. This approximation corresponds to a spinning particle or a pole-dipole interaction. Applying energy-momentum conservation and Einstein's field equations we obtain the Matthison-Papapetrou-Dixon (MPD) equations \cite{Mathisson:1937zz, Papapetrou:1951pa, Dixon:1970zz}
\begin{align}
&{\cal D}_s p^\lambda=-\frac{1}{2} t^\nu {\cal S}^{\rho\sigma} R^\lambda_{\;\;\nu\rho\sigma}\label{MPD1}\\
&{\cal D}_s {\cal S}^{\mu\nu}= p^\mu t^\nu-t^\mu p^\nu\,,\label{MPD2}
\end{align}
where $t^\mu$ is the normalized tangent vector.

Notice that the MPD system is not a closed set of differential equations and it must be suplemented with additional conditions. These conditions help to specify the reference worldline used to calculate the moments. It is tempting to think that one should simply choose the center of mass of the body but this is a rather subtle point in relativistic systems. Indeed, the notion of center of mass is observer-dependent, see \cite{Costa:2011zn} for an informative exposition. In practice, the choice of worldline is implemented by requiring that
\be \label{SC general}
{\cal S}^{\mu\nu}v_\mu=0\,,
\ee
for some normalized vector field $v^\mu$. The condition \eqref{SC general} identifies the solution of the MPD problem with the trajectory of the center of mass as measured in the rest frame of an observer moving with three-velocity\footnote{Recall we are working in 2+1 dimensions.} $v^\mu$. In the literature there are two widely used conditions, the Tulczyjew-Dixon (TD) condition 
\be \label{TD}
{\cal S}^{\mu\nu}p_\mu=0\,,
\ee
and the Mathisson-Pirani (MP) condition
\be \label{MP}
{\cal S}^{\mu\nu}t_\mu=0\,.
\ee
 Mathisson's helical motions arise if one chooses the MP condition to supplement the MPD system. As we shall demonstrate, the 2+1 dimensional  (MDP+MP) system corresponds to the shape equations \eqref{shape torsion2}. 

Since we are in 2+1 dimensions, the MP condition \eqref{MP} implies that the spin tensor can be written as 
\be \label{spin1}
{\cal S}^{\mu\nu}= \sigma\,\epsilon_{AB} \, n^{A\mu}n^{B\nu}\,,
\ee
where $\sigma$ is in constant due to \eqref{MPD1}. In turn, using Eq.\,\eqref{MPD2} we can write the momentum as 
\be\label{momentum 1}
\chi_t\, p^\mu=\left(t_\nu p^\nu\right) t^\mu-t_\nu{\cal D}_s{\cal S}^{\nu\mu}\,,
\ee
where $\chi_t= t_\mu t^\mu=\pm 1$ codifies the causal nature of the probe. Now, we show that $t_\nu p^\nu $ is actually constant. Indeed, equation \eqref{MPD1} and the properties of the Riemann tensor imply that $t_\mu{\cal D}_s p^\mu=0$. Moreover, we can show that 
\be
p^\mu{\cal D}_s t_\mu=-\chi_t \, t_\nu{\cal D}_s{\cal S}^{\nu\mu} {\cal D}_s t_{\mu}=\chi_t {\cal S}^{\nu\mu} \,{\cal D}_s  t_\nu {\cal D}_s t_{\mu}=0\,,
\ee
where the first equality follows from \eqref{momentum 1} and the parametrization by  arc-length, while the second equality comes from the MP condition. Thus, we have
\be\label{momentum}
-\chi_t\, p^\mu=\mu\, t^\mu+t_\nu{\cal D}_s{\cal S}^{\nu\mu}\,,
\ee
where $ \mu=-t_\nu p^\nu$ is constant along the curve and corresponds to the proper mass of the probe.

In order to relate the MPD+MP system to the shape equations it is convenient to rewrite \eqref{momentum} as 
\be\label{momentum 2}
-\chi_t\, p^\mu=\mu\, t^\mu+\sigma\,\epsilon_{AB} K^An^{B\mu}\,.
\ee
In passing, notice that 
\be\label{pp}
p^\mu p_\mu=\chi_t\left(\mu^2-\sigma^2 k_{\mathrm{FS}}^2\right)\,,
\ee
where we used that $\chi_t^2=1$ and $\det(\eta)=-\chi_t$ .
Furthermore, we can show that 
\be
\epsilon_{AB}{\cal D}_s\left(K^A n^{B\mu}\right)=-\epsilon_{AB} \, n^{A\mu}({\tilde D}K)^B\,.
\ee
The above equation together with the arc-length parametrization imply that the directional derivative of $ p^\mu$ is entirely normal. Explicitly, we have 
\be
n_{A\mu}{\cal D }_s(\chi_t\, p^\mu)=\mu\,K_A+\sigma \epsilon_{AB}({\tilde D}K)^B\,.
\ee
Therefore, in terms of extrinsic geometry the MPD equation \eqref{MPD1} reads
\be 
\mu\,  K_A  
+
\sigma \,  \epsilon_{AB}
\left[
(\tilde{D} K)^B
+
\chi_t R_s^{\;B}\right] = 0\,.\label{shape torsion MPD}
\ee
Finally, using $k^A=\tr K^A=\chi_t K^A$ we obtain 
\be 
\mu\,  k_A  
+
\sigma \,  \epsilon_{AB}
\left[
(\tilde{D} k)^B
+
R_s^{\;B}\right] = 0\,,\label{shape torsion MPD 1}
\ee
which matches perfectly the shape equation \eqref{shape torsion2} upon identifying $\mu \leftrightarrow \mathfrak{m}$ and $\sigma \leftrightarrow \mathfrak{s}$. There fore we have shown that, for an arbitrary three-dimensional manifold, the system of MPD+MP equations originates from a variational principle.

%
%
%

\subsection{Extrema in maximally symmetric spaces}\label{MPD}
\label{sec:MSS probes}

In maximally symmetric three-dimensional manifolds the Ricci tensor is proportional to the metric, which implies that \eqref{shape torsion2} simplifies to
\be \label{shape in MSS}
\mathfrak{m} k_A  
+
\mathfrak{s} \,  \epsilon_{AB}
(\tilde{D} k)^B= 0\,.
\ee
Clearly, geodesics are amongst the solutions to these equations.
 Moreover, equation \eqref{change in kFS} implies that any extremal curve in this kind of manifolds must have constant FS curvature. Furthermore,  equation \eqref{shape in MSS} implies that
\be
\kappa_{\mathrm{FS}}\left(\mathfrak{m}-\mathfrak{s}\,\tau_{\mathrm{FS}}\right)=0\,,
\ee
which implies that whenever the FS curvature and $\mathfrak{s}$ are non-vanishing the torsion is fixed to a particular constant
\be\label{tau MMS}
\tau_{\mathrm{FS}}=\frac{\mathfrak{m}}{\mathfrak{s}}\,.
\ee
Thus, extrema in maximally symmetric spaces correspond to curves with constant FS curvature and torsion. 

Now that we have figured out the behaviour of the extrinsic geometry we try to elucidate the shape of the curves themselves. We move to the FS gauge and by convention identify the normal index $A=1$ with the FS normal direction. Expanding equation \eqref{directional2} we find 
\begin{align}
&{\cal D}_s n^{(1)\mu}=\chi_{t}\,\kappa_{\mathrm{FS}}\,t^\mu-\eta_{22}\,\tau_{\mathrm{FS}}\,n^{(2)\mu} \label{FS eqs1}\\
&{\cal D}_s n^{(2)\mu}=\eta_{11}\,\tau_{\mathrm{FS}}\,n^{(1)\mu}\,,\label{FS eqs2}
\end{align}
where $\chi_{t}=t^\mu t_\mu$ encodes the causal nature of the probe. Notice that in the present convention $\eta_{11}=\chi_{\mathrm{FS}}$. Moreover, combining Eq.\,\eqref{normal} with \eqref{directional2} we obtain
\be
{\cal D}_s t^\mu\,=-\eta_{11}\,\kappa_{\mathrm{FS}}\,n^{(1)\mu}\,. \label{FS eqs3}
\ee
These are nothing but the FS equations for the moving frame. In a flat background we can reduce this system to a single differential equation for the curve $\gamma^\mu(s)$
\be\label{together}
\frac{d^4\gamma^\mu}{ds^4}+\left[\chi_t\,k^2_{\mathrm{FS}}+\det(\eta)\,\tau^2_{\mathrm{FS}}\right]\frac{d^2\gamma^\mu}{ds^2}=0\,.
\ee
If the ambient is Euclidean, this equation becomes simply
\be\label{Eq R3}
\frac{d^4\gamma^\mu}{ds^4}+\left(k^2_{\mathrm{FS}}+\tau^2_{\mathrm{FS}}\right)\frac{d^2\gamma^\mu}{ds^2}=0\,,
\ee
where $k^2_{\mathrm{FS}}$ is positive definite. On the other hand, if the ambient space has Lorentizan signature equation \eqref{together} reads
\be\label{Eq M3}
\frac{d^4\gamma^\mu}{ds^4}+\chi_t\left(k^2_{\mathrm{FS}}-\tau^2_{\mathrm{FS}}\right)\frac{d^2\gamma^\mu}{ds^2}=0\,,
\ee
where we used the fact that $\det(\eta)=-\chi_t$. In this case there are two signs  to be accounted for, each associated with the causality of the probe and that of its extrinsic curvature vector.

To find the analogue of \eqref{together} in a curved ambient space such as $\mathbb{H}_3$ or AdS$_3$ is less straightforward. The key difficulty lies on the fact that ${\cal D}_s$ no longer reduces to an ordinary derivative. To circumvent this complication it is convenient to regard the ambient space in question as a hypersurface embedded in a four-dimensional flat space. Indeed, both $\mathbb{H}_3$ and AdS$_3$ can be described as the zero set of $x^\alpha x_\alpha+L^2$ in $\mathbb{R}^4$,  the difference between the two being which metric is used to contract the $x^\alpha$: $\mathrm{diag}(-,+,+,+)$ for $\mathbb{H}_3$ and $\mathrm{diag}(-,-,+,+)$ for AdS$_3$. In this setting, the FS equations \,\eqref{FS eqs1}, \eqref{FS eqs2} and \eqref{FS eqs3} read
\begin{align}
&{\cal D}_s n^{(1)\alpha}=\chi_{t}\,\kappa_{\mathrm{FS}}\,t^\alpha-\eta_{22}\,\tau_{\mathrm{FS}}\,n^{(2)\alpha}\\
&{\cal D}_s n^{(2)\alpha}=\eta_{11}\,\tau_{\mathrm{FS}}\,n^{(1)\alpha}\,,\\
&{\cal D}_s t^\alpha\,=-\eta_{11}\,\kappa_{\mathrm{FS}}\,n^{(1)\alpha}\,,
\end{align}
and the directional derivative is given by the projection of the derivative along the submanifold
\be
{\cal D}_s\,V^\alpha = \partial_sV^\alpha + \frac{1}{L^2}\left(\gamma_\beta \partial_s V^\beta \right) \gamma^\alpha\,.
\ee
Combining the above results, we obtain the master equation
\be\label{master}
\frac{d^4\gamma^\alpha}{ds^4}+\left[\chi_t\,k^2_{\mathrm{FS}}+\det(\eta)\,\tau^2_{\mathrm{FS}}-\frac{\chi_t}{L^2}\right]\frac{d^2\gamma^\alpha}{ds^2}-\chi_t\det(\eta)\left(\frac{\tau_{\mathrm{FS}}}{L}\right)^2\gamma^\alpha=0\,,
\ee
where solutions ought to fullfill the constraint $\gamma^\alpha\gamma_\alpha=-L^2$.
For the hyperbolic space $\mathbb{H}_3$, equation \eqref{master} can be further reduced to 
\be\label{Eq H3}
\frac{d^4\gamma^\alpha}{ds^4}+\left(k^2_{\mathrm{FS}}+\tau^2_{\mathrm{FS}}-\frac{1}{L^2}\right)\frac{d^2\gamma^\alpha}{ds^2}-\left(\frac{\tau_{\mathrm{FS}}}{L}\right)^2\gamma^\alpha=0\,.
\ee
Meanwhile, for AdS$_3$ we have
\be\label{Eq AdS3}
\frac{d^4\gamma^\alpha}{ds^4}+\chi_t\left(k^2_{\mathrm{FS}}-\tau^2_{\mathrm{FS}}-\frac{1}{L^2}\right)\frac{d^2\gamma^\alpha}{ds^2}+\left(\frac{\tau_{\mathrm{FS}}}{L}\right)^2\gamma^\alpha=0\,,
\ee
where in a similar fashion to Minkowski it is necessary to keep track of the causal nature of the probe and its extrinsic curvature vector.

In the next sections we find the curves that solve these equations. Experience with $\mathbb{R}^3$  hints to the fact that these solutions must be some kind of helices. 
A helix in $\mathbb{R}^3$ is a curve whose tangent vector makes a constant angle with a predetermined direction known as the axis. A classic result by Lancret and Saint Venant is that a necessary and sufficient condition for a curve to be a helix is that the \emph{Lancret ratio}
\be \label{Lancret}
\tau_{\mathrm{FS}}/\kappa_{\mathrm{FS}}\,,
\ee
 is constant along the curve. Helices with constant FS torsion, and hence constant FS curvature, are known as \emph{cylindrical helices}. Thus we can summarize this section by saying that the centers of mass of spinning probes in maximally symmetric spaces move along cylindrical helices. However, we must first clarify what is meant by a cylindrical helix in spaces other than $\mathbb{R}^3$; this will be the focus of the next sections.

\subsection{ Mathisson's helical trajectories}

\label{sec:helical trajectories}

We start with the simplest non-trivial case, a timelike probe in Minkowski space. 
Cylindrical helices in Minkoswki space have been studied from a purely geometric perspective in \cite{2008arXiv0810.3351L}. In Section \ref{MPD}, we showed that these helical shapes must correspond to solutions of the Mathisson-Papapetrou-Dixon dipole equations \eqref{MPD1} and \eqref{MPD2} with the Mathisson-Pirani complementary condition \eqref{MP}. These trajectories are know as Mathisson's helical trajectories and surprinsingly their proper physical interpretation has been understoond only recently in \cite{Costa:2011zn, Costa:2017kdr}. In this brief aside we bridge between the geometrical language of the shape equations and the three-dimensional version of the insights presented in \cite{Costa:2011zn}.

In flat ambient geometries the shape of extrema are dictated by Eq.\,\eqref{together}, whose generic\footnote{By generic, we mean that $\tau_{\mathrm{FS}}\neq \kappa_{\mathrm{FS}}$.} solutions can be written as
\be\label{generic sol flat}
\gamma^\mu(s)=A^\mu+s\,B^\mu+\cos\left(\sqrt{\Lambda}s\right)\,C^\mu+\sin\left(\sqrt{\Lambda}s\right)\,D^\mu\,,
\ee
where 
\be\label{sol flat}
\Lambda=\chi_t\,k^2_{\mathrm{FS}}+\det(\eta)\,\tau^2_{\mathrm{FS}}\,,
\ee
and the coefficients in \eqref{generic sol flat} are subject to the condition $\dot\gamma_\mu \dot\gamma^\mu=\chi_t$.
In particular, for a timelike probe in Minkowski space we have
\be\label{sol mink time}
\Lambda=\tau^2_{\mathrm{FS}}-k^2_{\mathrm{FS}}\,,
\ee
with $k^2_{\mathrm{FS}}\geq 0$. Thus, unlike the Euclidean case (where $\Lambda\geq 0$ always) the Minkowski ambient allows for hyperbolic as well as trigonometric solutions. However, naively examining Mathisson's helical solutions presented in \cite{Costa:2011zn}, it seems that only trigonometric paths are to be considered. One might wonder whether there is something interesting hidding behind this simple observation.

The objective of the MPD equations is to provide an approximate description of the movement of an extended body in relativity by tracking the movement of a single point.  We encounter a similar problem in Newtonian physics and there we simply follow the trajectory of the center of mass. In a relativistic system, however, the notion of center of mass is less straightforward\footnote{We  recommend the reader to look at \cite{Costa:2011zn, Costa:2017kdr} for a very illuminating discussion.}. Indeed, the notion of center of mass is observer-dependent. In the case of the MPD system, it is the supplementary condition which determines which is the point to be tracked along the body trajectory. This choice entails the designation of a particular observer and the solutions of MPD delineate the trajectory of the center of mass as measured by that observer. We refer to these respectively as the TD observer or the MP observer depending on whether we impose \eqref{TD} or \eqref{MP}. The Lorentz factor between the MP and TD frames is given by 
\be
\frac{1}{\gamma^2}=\frac{{\cal M}^2 }{\mathfrak{m}^2}\,,
\ee
where
\be
 {\cal M}^2 =-p^\mu p_\mu\,,
\ee
is the proper mass observed by the TD observer. 
Using Eqs.\,\eqref{pp} and \eqref{tau MMS} we find
\be
\frac{1}{\gamma^2}=1-\left(\frac{k_{\mathrm{FS}}}{\tau_{\mathrm{FS}}}\right)^2\,.
\ee
Thus, we reach the conclusion that the Lancret ratio \eqref{Lancret} in Minkowski space corresponds to the relative speed between the MP and the TD frame. Requiring this speed to be subluminal implies that $\Lambda$ is positive and thus the solutions in \eqref{generic sol flat} are always trigonometric.

\section{Helices in hyperbolic and Anti de Sitter space}\label{helices}

\label{sec:solutions}

In this section we study helical trajectories in negatively curved spaces. We start off in the hyperbolic space $\mathbb{H}_3$, whose signature is Euclidean and then we explore AdS$_3$ where the subtleties associated with a Lorenzian signature must be considered. Extrema in $\mathbb{H}_3$ are curves of constant $k_\mathrm{FS}$ and $\tau_\mathrm{FS}$ satisfying \eqref{Eq H3}, solutions to the latter are of the form
\be\label{generic sol H3}
\gamma^\alpha(s)= \mathrm{M}^\alpha_{\;\;j} \mathrm{ v}^j\qquad
\mathrm{ v}=
\left( \begin{array}{c}
\cosh(\lambda s) \\
\sinh(\lambda s) \\
\cos(\omega s) \\
\sin(\omega s) \end{array} \right)\, ,
\ee
where
\be\label{freq H3}
\lambda=\sqrt{-\frac{\Lambda}{2}+\sqrt{\left(\frac{\Lambda}{2}\right)^2+\left(\frac{\tau_\mathrm{FS}}{L}\right)^2}}\qquad\omega=\sqrt{\frac{\Lambda}{2}+\sqrt{\left(\frac{\Lambda}{2}\right)^2+\left(\frac{\tau_\mathrm{FS}}{L}\right)^2}}\,,
\ee
and 
\be
\Lambda=k^2_{\mathrm{FS}}+\tau^2_{\mathrm{FS}}-\frac{1}{L^2}\,.
\ee
Notice that both $\lambda$ and $\omega$ in \eqref{freq H3} are real for any value of $k^2_{\mathrm{FS}}$, $\tau^2_{\mathrm{FS}}$ and $L$. Recall that $\gamma^\alpha$ in Eq.\,\eqref{generic sol H3} represents a curve in four dimensions and for it to lie on $\mathbb{H}_3$ it must satisfy the embedding condition $\gamma^\alpha\gamma_\alpha=-L^2$.  In addition to the restrictions beared by the embedding condition the constant coefficients in \eqref{generic sol H3} are further constrained by the tangent vector normalization condition $\dot \gamma^\alpha\dot \gamma_\alpha=1$.

\begin{center}
\begin{figure}
  \hspace{1cm}\includegraphics[width=.96\textwidth]{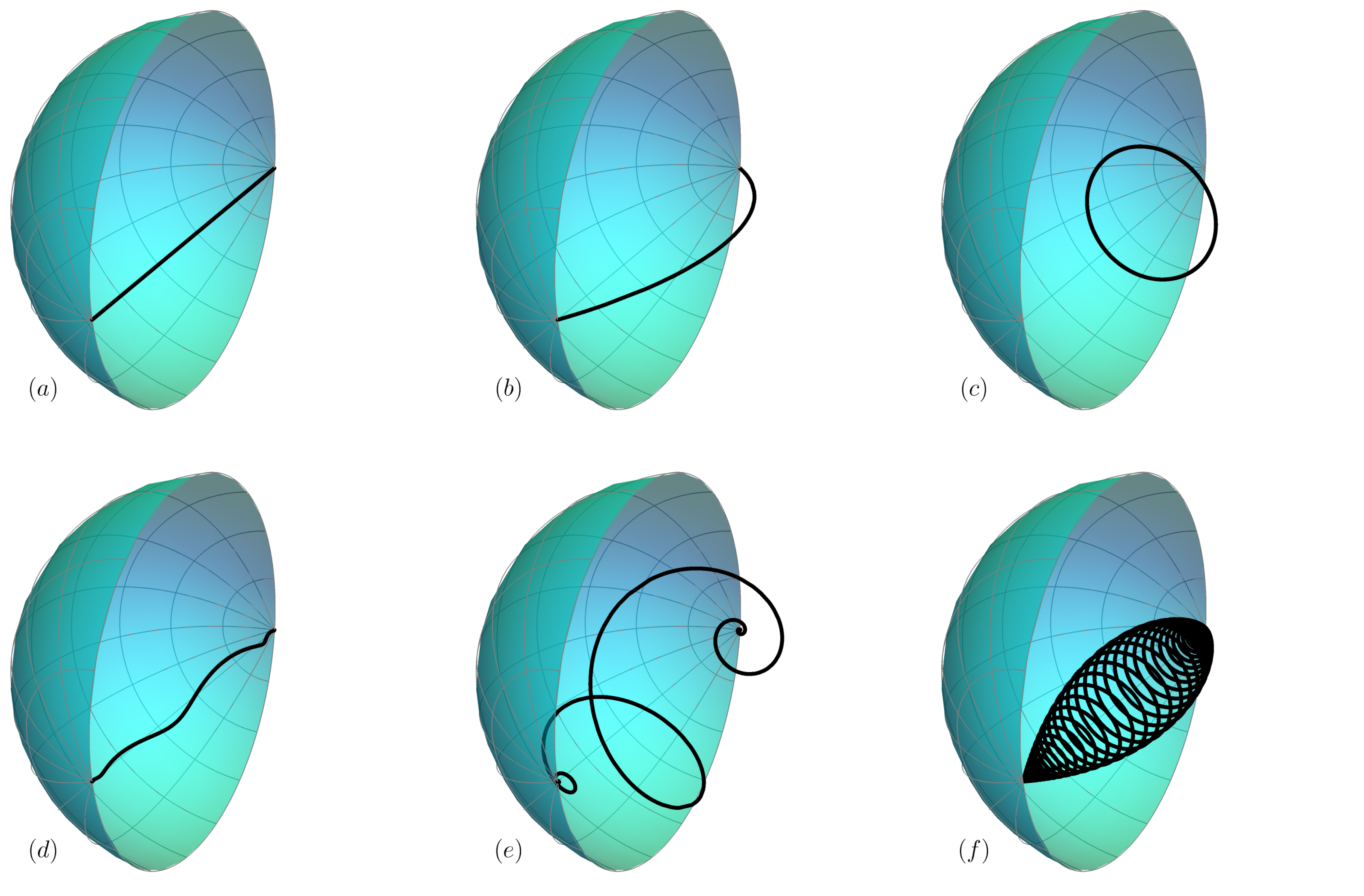}
  \caption{
Solutions $\tilde{\gamma}$ of equation \eqref{generic sol H3} obtained with \eqref{constr emb H3} and seed \eqref{H3seed}. For drawing purposes, we compactified $\mathbb{H}_3$ onto the Poincar\'e sphere of radius $L$, and constructed the projection so that if a curve ends on the boundary, its endpoints are antipodal. Only half of the sphere is displayed to show the interior.
\textbf{a)} Geodesic. 
\textbf{b)} Constant curvature curve with zero torsion and $L k_{FS}< 1$, the curve consists of an circle arc of radius bigger than $L$.
\textbf{c)} Constant curvature curve with zero torsion and $L k_{FS}> 1$, the curve consists of a full circle of radius smaller than $L$.
\textbf{d)} Helix with $L k_{FS} < 1$, as in case b), but with $L \tau_{FS} = 2$. The number of twists is infinite as the boundary is approached. 
\textbf{e)} Helix with $L k_{FS} > 1$ and $L \tau_{FS} = 1/10$. 
\textbf{f)} Helix with $L k_{FS}$ twice as of the one of e) and $L \tau_{FS} = 1/10$.
}
\label{H3curves}
\end{figure}
\end{center}

The embedding condition implies that the matrix of coefficients $\mathrm{M}$ in \eqref{generic sol H3} is not arbitrary. Indeed, in order for this requirement to be satisfied we must have
\be\label{constr emb H3}
\mathrm{M}^\top \eta_{(1,3)}\mathrm{M}=\mathrm{diag}(-a^2,a^2,b^2,b^2)\,,\qquad b^2-a^2=L^2\,.
\ee
Observe that every pair of solutions can be connected by means of an isometry i.e.~by left-multiplying the matrix $\mathrm{M}$ with a matrix 
${\cal I} \in O(1,3)$,  such that
\be\label{using isom}
{\cal I}^\top\eta_{(1,3)}{\cal I}=\eta_{(1,3)}\,.
\ee
It is therefore possible to construct every solution starting from a \emph{seed solution} $\tilde\gamma=\tilde{\mathrm{M}}\mathrm{v}$. For the present case a convenient choice of seed is generated by
\be
\tilde{\mathrm{M}}=\mathrm{diag}(a,a,b,b)\,.
\label{H3seed}
\ee
Finally, by demanding the normalization of the tangent vector $\dot \gamma^\alpha\dot \gamma_\alpha=1$ on the seed solution we find 
\be
(a\lambda)^2-(b\omega)^2=1\,.
\ee
The above condition is also invariant under the action of isometries, see Fig.~\ref{H3curves} .

\begin{center}
\begin{figure}
  \hspace{0cm}\includegraphics[width=.96\textwidth]{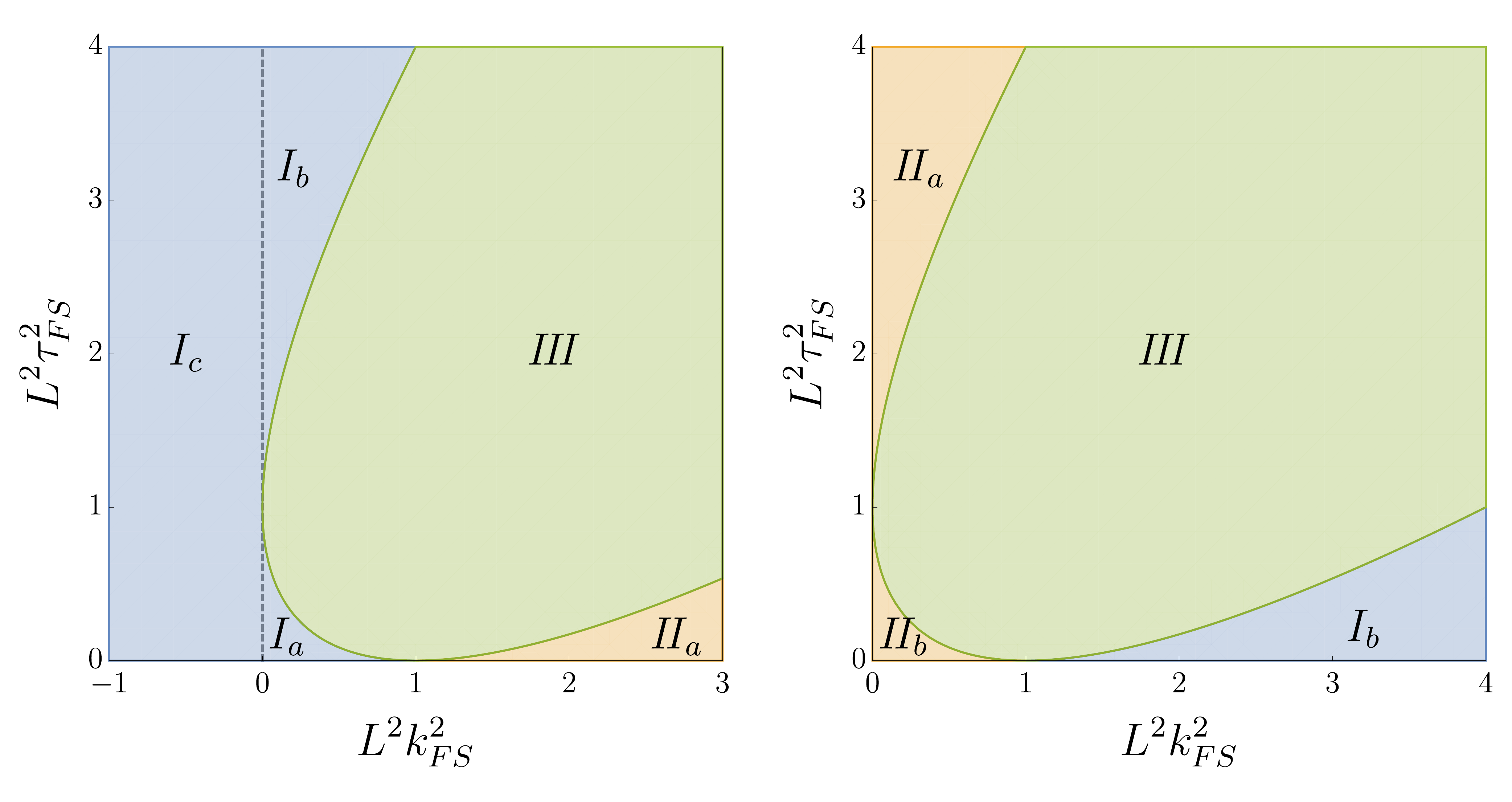}
  \caption{
Region plots showing  which case is pertinent for a given value of total curvature $k_{FS}^2$ and squared torsion $\tau_{FS}^2$ for curves in $AdS_3$. The left panel refers to space-like probes with $\chi_t=1$ (and hence both $\chi_{FS}$ and $k_{FS}^2$ can be negative). The right panel shows time-like probes with $\chi_t=-1$ and which can have only space-like FS normals. Blue regions refer to Case $I$, yellow to Case $I\!I$ and green to Case $I\!I\!I$. Notice however that different seeds (i.e. different permutations of $\mathrm{v}^j$) are needed to generate real solutions within a given region: for example the Case $I$ splits for space-like probes into three sub-regions, labeled $I_a$, $I_b$ and $I_c$ which correspond respectively to seeds  \eqref{cIa}, \eqref{cIb} and \eqref{cIc}.}
\label{Reg}
\end{figure}
\end{center}

A similar procedure can be followed to find the extrema corresponding to AdS$_3$ but some subtleties must be taken into account. To understand the solutions of Eq.\,\eqref{Eq AdS3} we must study carefully the behavior of the roots of the characteristic polynomial associated with this equation. The characteristic polynomial is biquadratic, and it s four roots are then of the form $\pm \lambda_\pm$ with 
\be
\lambda_\pm=\sqrt{z_\pm}\,,
\ee
where $z_{\pm}$ are zeros of the auxiliary polynomial 
\be\label{auxiliary}
z^2+\Lambda z+\left(\frac{\tau_{\mathrm{FS}}}{L}\right)^2\,,
\ee
with
\be
\Lambda=\chi_t\left(k^2_{\mathrm{FS}}-\tau^2_{\mathrm{FS}}-\frac{1}{L^2}\right)\,.
\ee
The nature of the roots is encoded in the discriminant
\begin{align}
&\label{discri2}\Delta=\Lambda^2-4 \left(\frac{\tau_{\mathrm{FS}}}{L}\right)^2\,.
\end{align}
Setting aside the degenerate cases ($\tau_{\mathrm{FS}}=0$ or $\Delta=0$) there are three possible scenarios:
\begin{itemize}
       \item \emph{Case} $I$: If $\Delta>0$ and $\Lambda<0$, then both $\lambda_\pm$ are real and distinct.
       \item \emph{Case} $I\!I$: If $\Delta>0$ and $\Lambda>0$, then both $\lambda_\pm$ are imaginary and distinct.
       \item \emph{Case} $I\!I\!I$: If $\Delta<0$, then both $\lambda_\pm$ are complex with non-zero imaginary and real parts, and $\lambda_+=\bar\lambda_-$.
\end{itemize}
The values of $k_{\mathrm{FS}}$ and $\tau_{\mathrm{FS}}$ determine to which case a solution belong, see Fig.\,\ref{Reg} for details.

\begin{figure}
  \includegraphics[width=.5\textwidth]{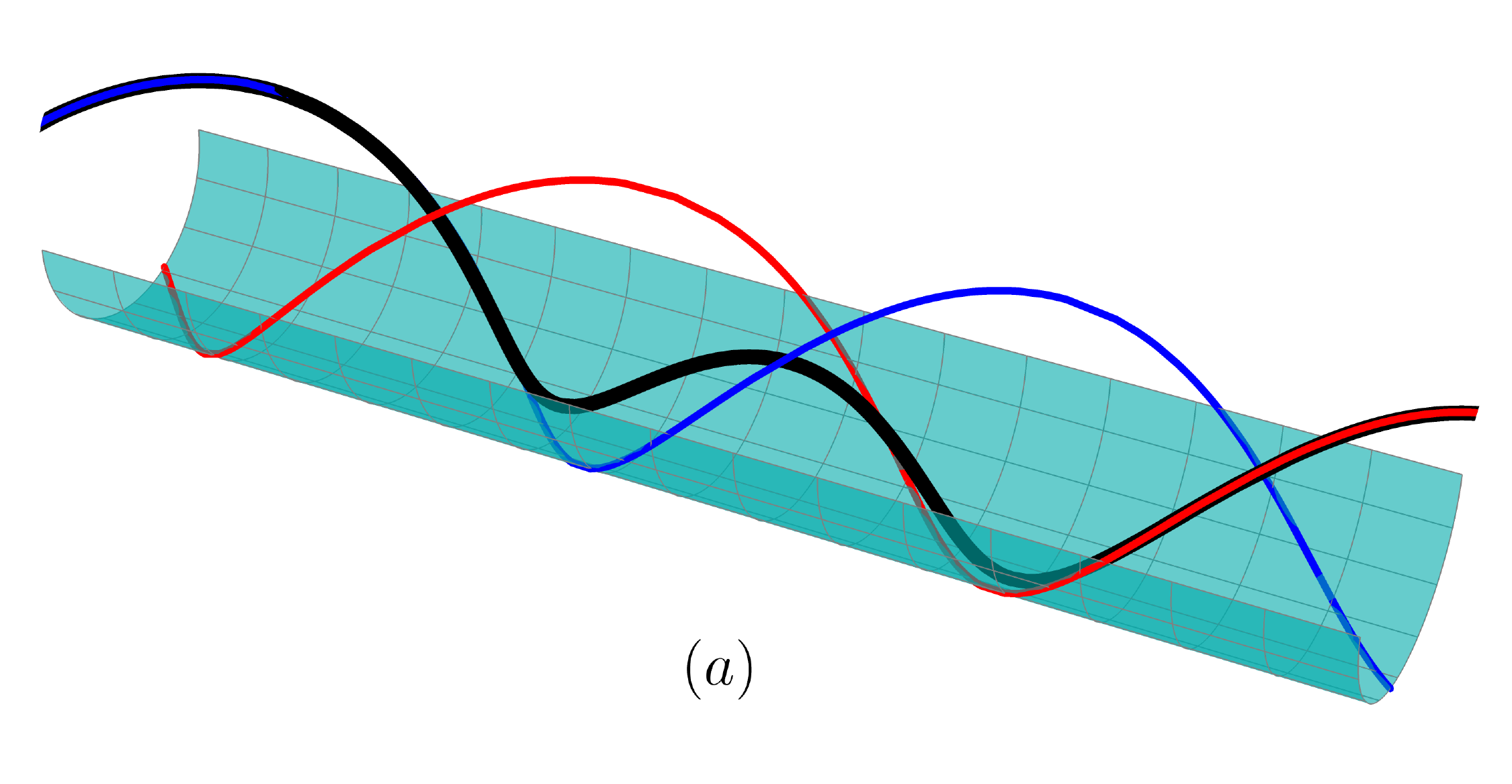}
  \includegraphics[width=.5\textwidth]{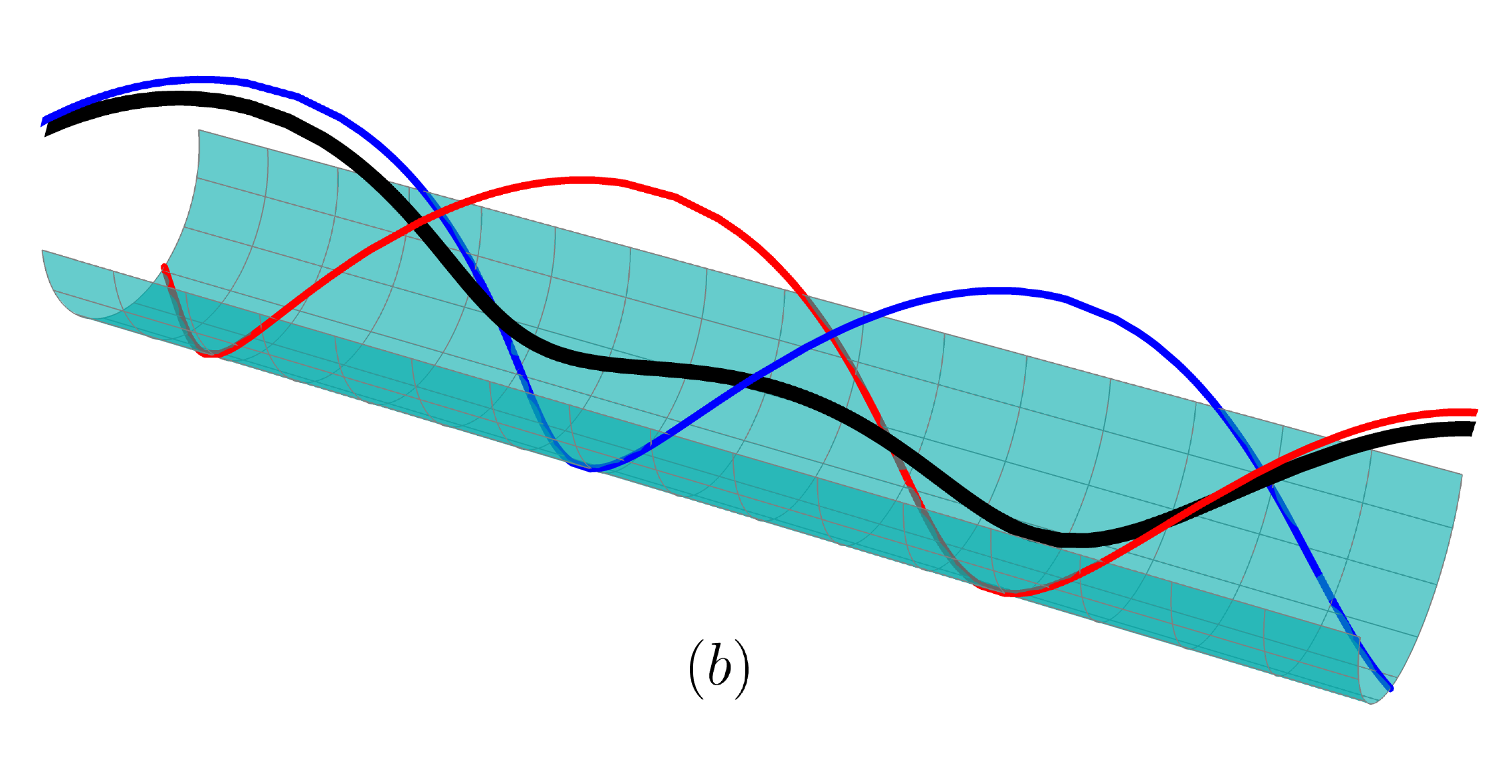}
\caption{Examples of helices belonging to case $I\!I\!I$. We display solutions embedded in the universal cover of $AdS_3$, i.e. the inside of a cylinder where the time coordinate spans along the axial direction. We depict respectively in red and blue the $s \to \pm \infty$ asymptotic null curves \eqref{asymptotic null}, towards which the solutions tend while approaching the conformal boundary.  We show only half of the boundary cylinder in light blue. \textbf{a)} Space-like helix with $L k_{FS} = L \tau_{FS} =3$.  \textbf{b)} Time-like helix with $L k_{FS} = L \tau_{FS} =1$. 
}
\label{figAdS3}
\end{figure}

We follow a similar strategy to the one used to determine helical motions in $\mathbb{H}_3$. Solutions to equation \eqref{Eq AdS3} for Case $I$ can be written as
\be\label{solution region1}
\gamma^\alpha(s)= \mathrm{M}^\alpha_{\;\;j} \mathrm{ v}^j
\qquad
 \mathrm{ v}=
\left( \begin{array}{c}
\cosh(\lambda_+ s) \\
\cosh(\lambda_- s) \\
\sinh(\lambda_+ s)  \\
\sinh(\lambda_- s)\end{array} \right)\,.
\ee
A natural choice would be to simply pick 
\be
\tilde{\mathrm{M}}=\mathrm{diag}(b,a,b,a)\,,
\ee
which upon requiring 
\be \label{cIc}
a^2+b^2=L^2\qquad (a\lambda_-)^2+(b\lambda_+)^2=\chi_t\,,
\ee
creates a seed which satifies the embedding and tangent vector normalization conditions.
However, this is not the whole picture since there is a priori no reason for choosing the ordering of the rows of $\mathrm{v}$ as in Eq.\,\eqref{solution region1}. This fact can be accounted for by considering shufflings ${\cal P}$ of the columns of $\tilde{\mathrm{M}}$. A careful consideration shows that (up to isometries) there are only two relevant permutations:
\begin{align}
&{\cal P}_1=\{1, 4, 3, 2\}\qquad b^2-a^2=L^2 \qquad (b\lambda_+)^2-(a\lambda_-)^2=\chi_t \label{cIa}\\
&{\cal P}_2=\{3, 2, 4, 1\}\qquad a^2-b^2=L^2 \qquad (a\lambda_-)^2-(b\lambda_+)^2=\chi_t\, , \label{cIb}
\end{align}
where we have also displayed the restrictions on the coefficients implied by the embedding and tangent vector normalization constraints. To stress the importance of these permutations, observe that for a spacelike probe Eq.\,\eqref{cIa} has real solutions only in the $k_{FS}^2\leq 0$ region of the $(k_{\mathrm{FS}}, \tau_{\mathrm{FS}})$-plane, which is only a subset of the region corresponding to Case $I$, see Fig.\,\ref{Reg}. Nevertheless, the whole region can be covered with (the closure of) regions where one and only one of the systems \eqref{cIa}, \eqref{cIb} or \eqref{cIc} has real roots, see Fig.\,\ref{Reg}. Timelike Case $I$ curves are simpler, in this case all the allowed region is covered by the seed $\tilde{\gamma}={\cal P}_2(\tilde{\mathrm{M}}) \mathrm{v}$. The upshot is that, for given values of $k_{\mathrm{FS}}$ and  $\tau_{\mathrm{FS}}$ falling in Case $I$, we must choose the right permutation of $\tilde{\mathrm{M}}$ to obtain real coefficients and then generate a seed accordingly. Once the seed has been determined, other solutions are then obtained via isometries of $\mathbb{R}^{(2,2)}$, see Figs.\,\ref{figAdS1}, \ref{figAdS2} and \ref{figAdS3} for illustrative examples.

\begin{figure}
  \hspace{.1cm}\includegraphics[width=\textwidth]{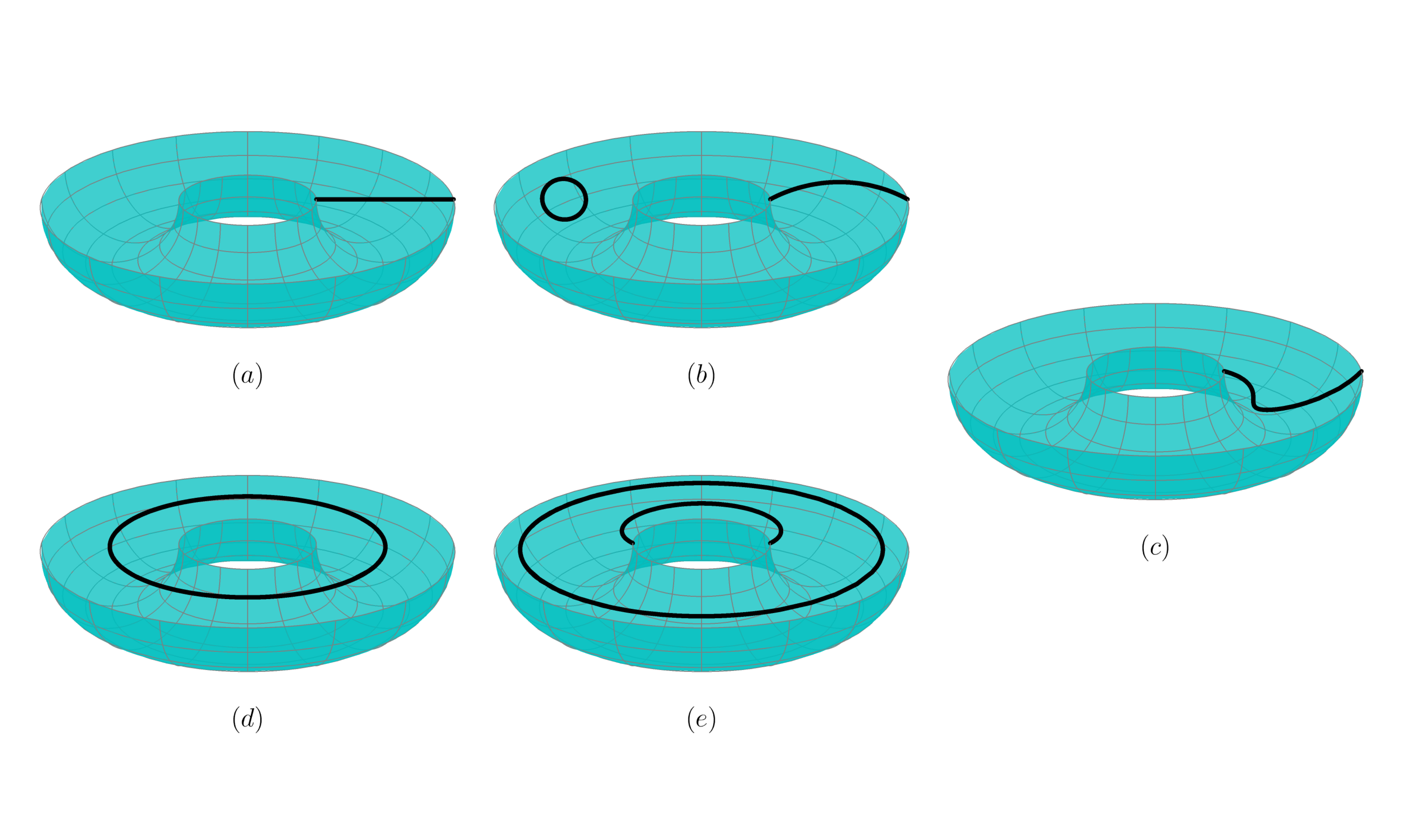}
  \caption{ 
Constant $k_{FS}$ curves in $AdS_3$ with $\tau_{FS}=0$. We represent global $AdS_3$ as a torus, since both boundary time and space are periodic (i.e. we do not take the universal covering of $AdS_3$). The $AdS_3$ boundary is depicted in light blue. As in Figure \ref{H3curves}, we display only half of the boundary in order to better show curves in the bulk. 
\textbf{a)} Spacelike geodesic, connecting two space-like separated boundary points. 
\textbf{b)} Spacelike curves with constant space-like curvature, with $L k_{FS}> 1$ (left circle) and $L k_{FS}< 1$ (right circle-arc), connecting two space-like separated boundary points. They belong repsectively to case $I\!I_a$ and $I_a$.
\textbf{c)} Spacelike curve with constant time-like curvature (i.e $\chi_{FS}=-1$) belonging to case $I_c$ and connecting two space-like separated boundary points.
\textbf{d)} Closed time-like geodesic. In $AdS$ spacetimes, time-like  geodesics can never connect two separate boundary points.
\textbf{e)} Time-like curves with constant space-like curvature, with $L k_{FS}< 1$ (larger full circle) and $L k_{FS}> 1$ (inner circle-arc), connecting two time-like separated boundary points. They belong repsectively to case $I\!I_b$ and $I_b$.
  }\label{figAdS1}
\end{figure}

In Case $I\!I$ all roots are purely imaginary and solutions of \eqref{Eq AdS3} take the form
\be
\gamma^\alpha(s)= \mathrm{M}^\alpha_{\;\;j} \mathrm{ v}^j
\qquad 
 \mathrm{ v}=\left( \begin{array}{c}
\cos(\omega_+ s) \\
\sin(\omega_+ s) \\
\cos(\omega_- s) \\
\sin(\omega_- s) \end{array} \right)\,,
\ee
where $\omega_\pm=-i\lambda_\pm$. Once more,  $\mathrm{M}$ must satisfy the embedding condition and we can generate a seed using
\be
\tilde{\mathrm{M}}=\mathrm{diag}(b,b,a,a)\,,
\label{cIIa}
\ee
where
\be
b^2-a^2=L^2\qquad (a\omega_-)^2-(b\omega_+)^2=\chi_t\,.
\ee
Just as in the previous case, it is important to consider the permutations of the columns of $\tilde{\mathrm{ M}}$. In the present case, it is the timelike probes that require more than one permutation. Indeed, the allowed region for Case $I\!I$ spacelike probes, see Fig.\,\ref{Reg}, is completely covered by the seed $\tilde{\mathrm{ M}}$. On the other hand, for a timelike probe we must consider, besides $\tilde{\mathrm{ M}}$, the additional permutation 
\begin{align}
{\cal P}_1=\{3, 4, 1, 2\}\qquad b^2-a^2=L^2 \qquad (a\omega_+)^2-(b\omega_-)^2=\chi_t\, , \label{cIIb}
\end{align}
to cover the whole region, see Fig.\,\ref{Reg}. As usual, once a seed is determined we can generate other solutions via isometries, see Figs.\,\ref{figAdS1} and \ref{figAdS2}.

\begin{figure}
  \hspace{0cm}\includegraphics[width=\textwidth]{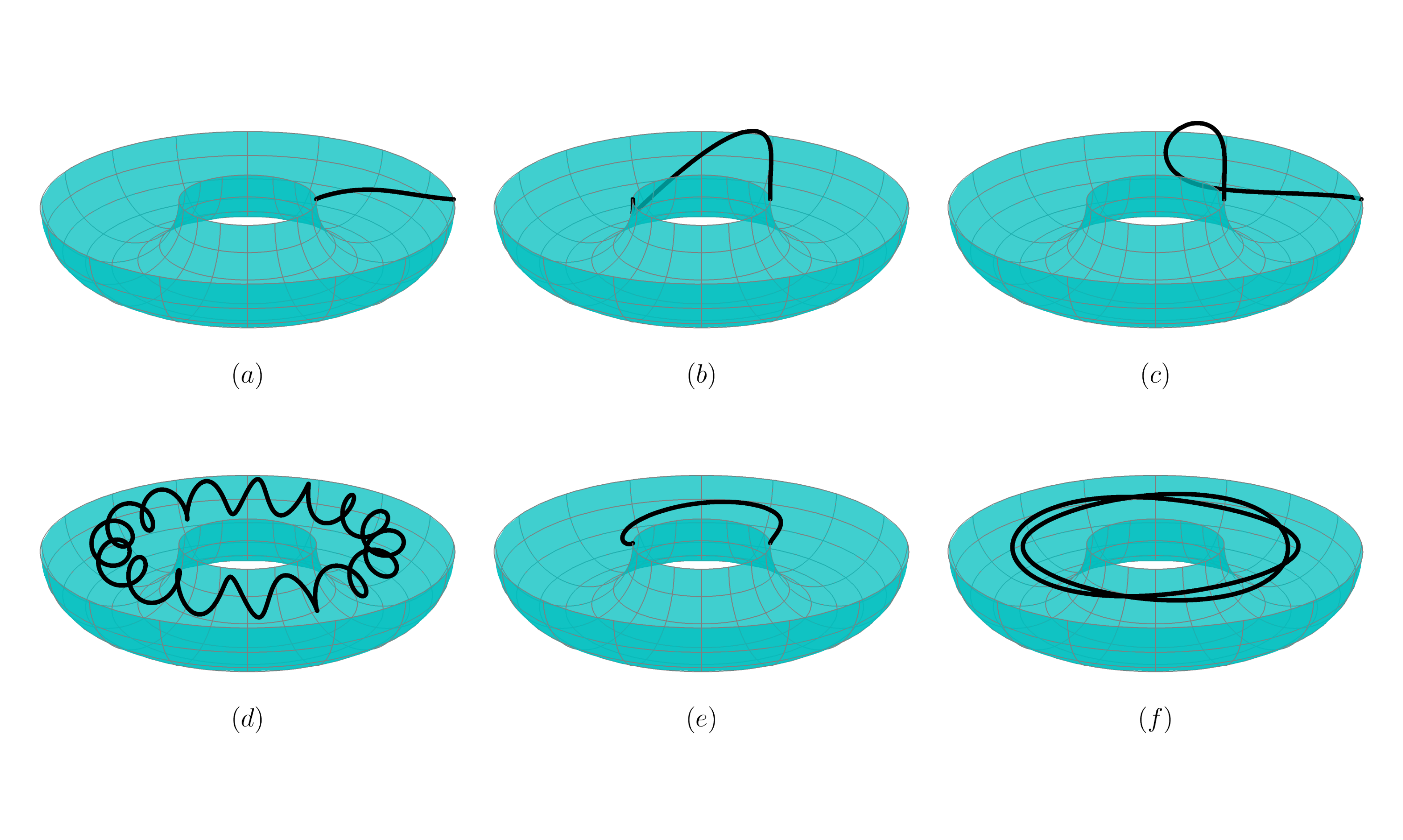}
\caption{
Continuation of Figure \ref{figAdS1}, showing curves with non-zero Frenet-Serret torsion $\tau_{FS}$.
\textbf{a)} Space-like helix belonging to case $I_a$  with $L k_{FS} = 1/4$ and torsion $L \tau_{FS} = 1/4$.
\textbf{b)} Space-like helix belonging to case $I_b$  with $L k_{FS} = 1/2$ and torsion $L \tau_{FS} = 2$ and, remarkably, connecting two time-like separated boundary points.
\textbf{c)} Space-like helix belonging to case $I_c$  with $\chi_{FS}=-1$, $L \kappa_{FS} = 1/4$ and torsion $L \tau_{FS} = 2$.
\textbf{d)} Space-like helix belonging to case $I\!I_a$  with $L k_{FS} = 2$ and torsion such that $\omega_+/\omega_-=16$.
\textbf{e)} Time-like helix belonging to case $I_b$  with $L k_{FS} = 2$ and torsion $L \tau_{FS} = 1/2$, ending on time-like separated boundary points, as in Figure \ref{figAdS1}.e.
\textbf{f)} Time-like helix belonging to case $I\!I_b$  with $L k_{FS} = 1/2$ and torsion such that $\omega_+/\omega_-=2$.
}\label{figAdS2}
\end{figure}

Finally, solutions for Case $I\!I\!I$ read
\be
\gamma^\alpha(s)= \mathrm{M}^\alpha_{\;\;j} \mathrm{ v}^j
\qquad 
 \mathrm{ v}=
\left( \begin{array}{c}
\cosh(\lambda s)\cos(\omega s) \\
\cosh(\lambda s) \sin(\omega s)\\
\sinh(\lambda s)\cos(\omega s) \\
\sinh(\lambda s)\sin(\omega s) \end{array} \right)\,,
\ee
were
\be
\lambda=\frac{1}{2}(\lambda_++\lambda_-)\qquad  \omega=\frac{1}{2i}(\lambda_+ -\lambda_-)\,.
\ee
The embedding and tangent vector normalization constraints can be solved using the matrix
\be
\tilde M=
\left( \begin{array}{cccc}
-b& 0& 0 & a \\
0&-b &  -a&0 \\
0& -a & b &0 \\
a&0& 0&b\end{array} \right)\,\,
\ee
with
\be\label{cIII}
b^2-a^2=L^2\qquad L^2(\omega^2-\lambda^2)-4ab\lambda\omega=\chi_t\,.
\ee
This system has real roots on all the Case $I\!I\!I$ region of the $(k_{\mathrm{FS}}, \tau_{\mathrm{FS}})$-plane and all seeds generated by other permutations with real roots are connected by isometries.

\subsection{Asymptotic behaviour}
\label{sec:asymptotic}

Now that the landscape of cylindrical helices in hyperbolic spaces has been thorougly explored, one might still wonder whether there is a more intuitive way for understanding the different types of solutions we encountered so far. It turns out that the various helices constructed above have very different asymptotic behaviour. We refer to the Appendix \ref{projs} for the notation of our coordinate representations of $\mathbb{H}_3$ and global $AdS_3$.

First, let us focus on the helices in $\mathbb{H}_3$. From \eqref{generic sol H3} and \eqref{H3 coord} one computes the radial profile of the solutions to be
\be
\left( \tanh \rho(s)/2 \right)^2 =  1 - \frac{2 L }{L + a \cosh s \lambda} \,,
\ee
which implies that $\rho \to \infty$ for $s \to \pm \infty$ if $\lambda\neq 0$. This means that all cylindrical helices in $\mathbb{H}_3$ reach the conformal boundary, unless the parameter $\lambda$ is equal to zero. By inspection of \eqref{freq H3}, it is easy to show that the latter condition holds only if
\be
\tau_{FS}=0 \quad \mathrm{and} \quad L k_{FS}\geq 1,
\ee
i.e. only if the curve is a planar with constant curvature, as the circle depicted in Figure \ref{H3curves}.c.

In a similar fashion, we are able to determine the asymptotic behaviour of helices in $AdS_3$, case by case. For case $I_a$, using the seed \eqref{cIa} one gets
\be
\left( \tanh \rho(s)/2 \right)^2  = 1 - \frac{2 L}{L + \sqrt{a^2 \sinh^2 \lambda_- s + b^2 \cosh^2 \lambda_+ s}} \,,
\label{rhoIa}
\ee
which shows how the curves will asymptotically approach the conformal boundary unless both roots $\lambda_\pm$ vanish simultaneously. It can be shown that for $\lambda_\pm=0$ to be satisfied, a curve needs to have $L^2 k_{FS}^2=1$ and $\tau_{FS}=0$, which corresponds to a degenerate solution where every point of the curve lies at the conformal boundary for every $s$. Discarding this degenerate case, it is possible to show how curves of type $I_a$ approach a specific point on the boundary given by
\be 
t(s) \xrightarrow[s\to \pm \infty]{} 0 \,,\;\; \phi(s) \xrightarrow[s\to \pm \infty]{} \pm \frac{\pi}{2} \,,
\label{endpointsIa}
\ee
which implies that the endpoints for seed \eqref{cIa} are space-like separated. 
Cases $I_b$ and $I_c$ produce a radial profile very similar to \eqref{rhoIa}, and thus also these curves end on the boundary. While case $I_c$ with seed \eqref{cIc} has exactly the same endpoints as $I_a$, i.e. \eqref{endpointsIa}, case $I_b$  has endpoints which are time-like separated
\be 
t(s) \xrightarrow[s\to \pm \infty]{} \{0,\pi\} \,,\;\; \phi(s) \xrightarrow[s\to \pm \infty]{} \frac{\pi}{2} \,.
\label{endpointsIb}
\ee
Note that this result is independent of the causal nature of the probe: therefore, curves with $\chi_t=1$ and belonging to case $I_b$ are space-like curves connecting time-like separated boundary points.

Curves belonging to case $I\!I$ are simpler, since it is easy to prove that all solutions lie at constant radial coordinate
\be
\left( \tanh \rho(s)/2 \right)^2  = 1 - \frac{2 L}{|b|+L} < 1 \,,
\ee
where the inequality follows from either relation \eqref{cIIa} or \eqref{cIIb}, both of which imply $|b|>L$. Therefore, all curves in Cases $I\!I_a$ and $I\!I_b$ never reach the boundary, as shown in the examples of Figures \ref{figAdS1} and \ref{figAdS2}.

Finally, curves belonging to case $I\!I\!I$ display a similar behaviour to case $I$, where the radial profile from seed \eqref{cIII} is 
\be
\left( \tanh \rho(s)/2 \right)^2  = 1 - \frac{4 L}{2L + \sqrt{2} \sqrt{L^2 + (a^2+b^2) \cosh 2\lambda s}} \,,
\ee
which shows that also these curves always end on the conformal boundary for $s \to \pm \infty$ (the case $\lambda=0$ reduces to the same degenerate curve of Case $I$). However, we find that instead of ending on fixed endpoints, these helices tend towards \textit{asymptotic null boundary curves}, defined by the relation
\be
\quad \cos^2 t(s) = \sin^2 \phi(s) \,,
\label{asymptotic null}
\ee
which implies $\phi(s) = \pm t(s) \pm \frac{\pi}{2}$. In Figure \ref{figAdS3} we represent these two asymptotic curves as a blue and a red helices of opposite chirality.

\section{Conclusions and discussion}

In this work we have explored the space of extrema of the functional \eqref{torsionaction1}. When the embedding manifold is a hyperbolic space form, this functional is of interest in computations of holographic entanglement entropy associated to anomalous CFTs \cite{Castro:2014tta}. If instead the curve is embedded in $\mathbb{R}^3$, then it can be interpreted as a torsion-dependent energy associated to a linear elastic structure (as it might be a long polymer) \cite{doi:10.1093/qjmam/hbn012}\cite{1751-8121-47-35-355201}.

In Section \ref{sec:shape equations} we derive the equations of motion associated to \eqref{torsionaction1} for an arbitrary embedding manifold, see \eqref{shape torsion2}. For maximally symmetric spaces (as $\mathbb{H}_3$ and $AdS_3$) such \textit{shape equations} take the very simple form \eqref{shape in MSS}. By choosing an appropriate local rotation acting on the normal bundle of the curve, we can express the shape equations in terms of Frenet-Serret frame, which makes evident how extrema of \eqref{torsionaction1} are \textit{cylidrical helices}, i.e. curves with constant total squared curvature and  Frenet-Serret torsion. While the value of the total curvature can be arbitrary, the torsion is fixed to be $\tau_{FS}=\mathfrak{m/s}$.

The shape equations are intimately related to the motion of spinning extended objects in curved space-times. Indeed, we show in Section \ref{sec:spinning particles} that the Mathisson-Papapetrou-Dixon (MPD) dipole equations \eqref{MPD1} and \eqref{MPD2} (see \cite{Mathisson:1937zz, Papapetrou:1951pa, Dixon:1970zz}) are exactly equivalent to \eqref{shape torsion2}, once the Mathisson-Pirani (MP) supplementary condition \eqref{MP} is implemented. The MPD equations have been extensively studied  for four-dimensional trajectories in Lorenzian manifolds, while much less has been said about the three-dimensional case. We have here proved that the MPD+MP system of equations can be obtained from a variational principle for an arbitrary gravitational background.

In the theory of relativistic spinning bodies it is known that the notion of center of mass is observer-dependent \cite{Costa:2011zn}. It is in fact the role of the spin supplementary condition to pick a specific observer. In the literature, besides the MP condition often the Tulczyjew-Dixon (TD) \eqref{TD} is also often implemented. We proved n Section \ref{sec:helical trajectories} that in our geometric language the Lancret ratio (i.e. the quantity  $k_{FS}/\tau_{FS}$) of an helix embedded in Minkowksy space is directly linked to the relative velocity between the MP and TD reference frames.

Once established that the extrema of \eqref{torsionaction1} are cylindrical helices, we explain in Section \ref{sec:MSS probes} how to find the actual embedding functions. For maximally symmetric space-times it turns out to be extremely convenient to view curves as embedded in four-dimensional Minkowsky space and constrained to move on a three-dimensional hyperbolic submanifold. In this way it is immediate to view the curve as a solution of a single master equation \eqref{master} which depends on the causal nature of both the probe (encoded by $\chi_t = \pm 1$, the norm of the curve's tangent vector) and of the Frenet-Serret normal (encoded by $\chi_{FS}$, the norm of the FS normal vector). Namely, while time-like curves always have a space-like FS normal, a space-like helix can have either a time-like or a space-like FS normal.

In Section \ref{sec:solutions} we explicitly solve the master equation for both $\mathbb{H}_3$ and $AdS_3$ spaces. While the former case is straightforward (see Figure \ref{H3curves}), the case for Anti-de Sitter contains several subtleties. As explained in Figure \ref{Reg}, solutions of different nature are possible depending on the values of $k_{FS}$ and $\tau_{FS}$ as well as of $\chi_t$ and $\chi_{FS}$. We recognize that helices in $AdS_3$ belong to three different classes, which eventually splits in a total of six different sub-classes. These different cases can be understood in terms of their asymptotic behaviour (see Section \ref{sec:asymptotic}):
\begin{itemize}
\item Case $I$ solutions always end on the conformal boundary. Curves of type $I_a$ and $I_c$ always end on space-like separated endpoints, while curves of type $I_b$ end always on time-like separated endpoints, regardless of the causal nature of the curve itself (see Figure \ref{figAdS1}). These  curves can be seen as deformations of $AdS_3$  geodesics and curves of constant curvature with $k_{FS}<1/L$.
\item Case $I\!I$ solutions lie always at a fixed radial distance, and therefore never reach the boundary. They generalize curves of constant curvature with $\kappa_{FS}$ larger than the $AdS_3$ radius.
\item Case $I\!I\!I$ solutions do also reach always the conformal boundary, but instead of ending on a fixed point, they asymptotize towards \textit{boundary null curves}: they interpolate between two boundary helices of opposite chirality, see Figure \ref{figAdS3}. Such curves have no analogue in Riemannian spaces.
\end{itemize}

There are a number of avenues for future reseach that follow naturally from the present considerations. First of all, it ought to be straightforward to map our helices to any Ba\~nados geometry, such as shockwaves or BTZ black holes. While local properties should remain intact, there might be interesting findings to be made from a global perspective. Moreover, the true importance of \emph{torsionfull} curves in the study of entanglement entropy emerges when geodesics are not amongst the extrema of \ref{torsionaction1}. This turs out to be the case for domain walls which are used to model renormalization group flows holographically. We have delved into this subject in an article which will be released shortly \cite{Us1}. Following \cite{Castro:2014tta}, we know that the solutions of the shape equations are a proxy for entanglement entropy in theories such as Topologically Massive Gravity (TMG) \cite{Deser:1982vy}. Interestingly, TMG admits non-AdS vacuum solutions known as warped AdS, these are spacetimes with non-vanishing Cotton tensor. Warped AdS geometries can be dealt with in a manner analogous to the embedding formalism used to construct the $\mathbb{H}_3$ and AdS$_3$ helices in Sec.\,\ref{sec:solutions}, see \cite{Bengtsson:2005zj}. We deem the construction of helices in warped AdS an interesting and physically relevant question and we are currently engaged in it.

\section*{Acknowledgements}

 PF is supported by The Netherlands Organization for Scientific Research. 
The work of AVO is supported by the NCN grant 2012/06/A/ST2/00396. AVO wishes to thank the theory group at CERN and the Yukawa Institute for Theoretical Physics for their hospitality during the development of this work. 
It is a pleasure to acknowledge Michael Abbott, Pawe\l{} Caputa, Alejandra Castro, Filipe Costa, Mario Flory and Jos\'e Natario for enlightening conversations and correspondence, as well as for helpful comments on earlier versions of this work. Especially, we wish to express our gratitute to Alejandra Castro for suggesting us to look into these questions. 

\newpage
\appendix
\section{Projections}\label{projs}

In this appendix we elaborate on the coordinate systems used to depict and study helices in the hyperbolic and Anti de Sitter Space. We first consider these spaces as hypersurfaces embedded in a four-dimensional flat space. For instance, the hyperbolic space can be described as the hypersurface in $\mathbb{R}^4$ with metric $\mathrm{diag}(-1,1,1,1)$ satisfying 
\begin{equation}
x^\alpha x_\alpha+L^2 = 0.
\end{equation}
One way to solve the equation of this hyperboloid is by 
\begin{align}\label{H3 coord}
x^1 = L \cosh \rho,\quad & x^2 = L \sinh\rho \cos \theta \cos\phi,\\ \nonumber \quad x^3 = L \sinh\rho \cos\theta\sin \phi, \quad& x^4 = L \sinh\rho\sin\theta
\end{align}
but in order to study the asymptotic behavior of the curves in $\mathbb{H}^3$, we must compactify the hyperbolic space. To do so, we introduce the finite variable
\begin{equation}
\label{compactrho}
\tilde \rho = \tanh \frac{\rho}{2}.
\end{equation}
The resulting coordinate system, corresponds to the Poincar\'e sphere of radius $L$.

Meanwhile, we consider $AdS_3$ to be the hypersurface
\begin{equation}
x^\alpha x_\alpha+L^2 = 0,
\end{equation}
in $\mathbb{R}^4$ with metric $\mathrm{diag}(-1,-1,1,1)$. One solution for this equation is given by the global coordinates of $AdS_3$:
\begin{equation}
x^1 = L \cosh\rho \cos t, \quad x^2 = L \cosh \rho \sin t, \quad x^3 = L \sinh \rho \sin \phi, \quad x^4 = L \sinh \rho \cos \phi.
\end{equation}
As in the case of $\mathbb{H}_3$, it is convenient to work in a compactified space, to make clear the behavior of the curves as they approach the boundary. For that purpose, we make use again of equation \ref{compactrho}.

Finally, for drawing purposes we represent $AdS_3$ as a torus. This representation is useful because the torus clearly shows the periodic behavior of the solutions described in section \ref{sec:solutions} for the case $II$. The function we use to map $AdS_3$ onto the torus with mayor radius $R$ and minor radius $L$ is given by
\begin{equation}
x = \left(R+ L \tilde \rho \sin \phi \right)\sin t, \quad y =  \left(R+ L \tilde \rho \sin \phi \right)\cos t, \quad z = L \tilde \rho \cos \phi.
\end{equation}

\newpage

\bibliography{biblio}{}

\end{document}